\magnification=1200
\hoffset=.0cm
\voffset=.0cm
\baselineskip=.53cm plus .53mm minus .53mm

%
%
%
%
\def\ref#1{\lbrack#1\rbrack}
%
%
%
%
\input amssym.def
\input amssym.tex
%
%
\font\teneusm=eusm10                    
\font\seveneusm=eusm7                   
\font\fiveeusm=eusm5                 
%
%

%
%

%
%
\newfam\eusmfam
\textfont\eusmfam=\teneusm
\scriptfont\eusmfam=\seveneusm
\scriptscriptfont\eusmfam=\fiveeusm

\def\proclaim #1. #2\par{\medbreak{\bf #1.\enspace}{\it #2}\par\medbreak}
%
%
%
%
%

\def\hst1{\hskip 1pt}

%
%
%
%
%

\def\CN{{\cal N}}

\hbox to 16.5 truecm{March 1999   \hfil DFUB 99--5}
\hbox to 16.5 truecm{Version 1  \hfil hep-th/9903161}
\vskip2cm
\centerline{\bf BOSONIC QUADRATIC ACTIONS FOR 11D}   
\centerline{\bf SUPERGRAVITY ON ${\rm AdS}_{7/4}\times {\rm S}_{4/7}$}
\vskip1cm
\centerline{by}
\vskip.5cm
\centerline{{\bf Fiorenzo Bastianelli}  and {\bf Roberto Zucchini}}
\centerline{\it Dipartimento di Fisica, Universit\`a degli Studi di Bologna}
\centerline{\it V. Irnerio 46, I-40126 Bologna, Italy}
\centerline{and}
\centerline{\it I. N. F. N., Sezione di Bologna}
\vskip1cm
\centerline{\bf Abstract} 
We determine from $11d$ supergravity the quadratic bulk action for the 
physical bosonic fields relevant for the computation
of correlation functions of normalized chiral operators in $D=6$, 
$\CN=(0,2)$ and $D=3$, $\CN=8$ supersymmetric CFT in the large $N$ limit,
as dictated by the AdS/CFT duality conjecture.
\par\noindent
\vfill\eject
\item{\bf 0.} {\bf Introduction}
\vskip.4cm
\par
The duality [1--3] between string/M theory on Anti-de Sitter space (AdS) 
times a compact manifold and conformal field theory (CFT) living on the 
boundary of AdS is quite useful in the program of extracting non-trivial 
information for various strongly coupled quantum field theories.
The most studied duality relates type IIB superstring theory on 
${\rm AdS}_5\times {\rm S}_5$ 
to  $D=4$, $\CN=4$ ${\rm SU}(N)$ super Yang-Mills (YM) 
theory. In the large $N$ limit, the AdS side of the duality 
can be approximated by type IIB supergravity. 
In one of the many applications [4], the latter has been employed to compute 
2 and 3 point functions for chiral primary operators of the  super YM theory, 
whose conformal dimensions had already been identified in [3] by relating 
them to the masses of the corresponding Kaluza--Klein scalars
on ${\rm AdS}_5\times {\rm S}_5$.
Various efforts are being made also to investigate the structure and compute
the more complex 4 point functions (see  [5--8] and references therein).

However, other interesting dualities with maximal supersymmetry 
(i.e. 16 supersymmetry charges on the CFT side) have been proposed in [1].
According to this proposal, M--theory on 
${\rm AdS}_7\times {\rm S}_4$ is dual to the 
$D=6$, $\CN=(0,2)$ supersymmetric CFT (to be denoted by SCFT$_6$).
Similarly, M--theory on 
${\rm AdS}_4\times {\rm S}_7$ is dual to the $D=3$, $\CN=8$  
supersymmetric CFT (to be denoted by SCFT$_3$).
These SCFTs are realized, for example, on the world-volume of $N$ coincident 
M5 and M2 branes, respectively. Yet, little is known about them,
and the AdS/CFT duality offers a useful tool to learn more about these
remarkable theories. Again, in the large $N$ limit, one can approximate 
M--theory by its low energy limit, $11d$ supergravity.
By using that, some groups have identified the conformal dimensions of the 
corresponding chiral operators [9--12]. Quite recently
the calculation of 3 point functions of chiral operators in the 
large $N$ limit of the SCFT$_6$ theory has appeared in [13]. 

In this paper, we determine the $11d$ supergravity quadratic bulk action of 
the physical bosonic fields relevant for the computation
of the correlation functions of normalized chiral operators of both 
SCFT$_6$ and SCFT$_3$. In particular, we 
carefully compute the normalization constants of the kinetic terms of the
scalar fields, as these numbers are a crucial input in such calculations.

We have found that the method originally described by Lee, Minwalla, 
Rangamani and Seiberg in [4] does not yield any sensible result, though we
have verified the correctness of their final answer. This fact has
recently been pointed out also Corrado, Florea and McNees [13]. 
Therefore, we carry out our analysis by extending the method of 
Arutyunov and Frolov [14], 
originally worked out for type IIB $10d$ supergravity on 
${\rm AdS}_5\times {\rm S}_5$, to $11d$ supergravity on  
${\rm AdS}_7\times {\rm S}_4$ and ${\rm AdS}_4\times {\rm S}_7$.
Contrary to what claimed in [13], we find that this approach  provides
the correct quadratic action. The method is completely self contained
in the sense that it does not require the knowledge of the action beyond
the quadratic approximation. As 
it consists in isolating the scalar fields of interest by simply 
performing field redefinitions, 
one does not need to put on-shell the scalar fluctuations
nor study complicated boundary terms in the supergravity action.
Moreover, it does not lead to either non local or 
higher derivative terms in AdS space. 

From our final results, one can read off the masses of the scalar 
fluctuations together with the searched for normalization factors. 
Needless to say that we reproduce the values of the masses
as originally worked out in [15] and in [16,17] for the  
${\rm AdS}_7\times {\rm S}_4$ and 
 ${\rm AdS}_4\times {\rm S}_7$ cases, respectively.
As for the normalization factors, we confirm the result recently
obtained for the ${\rm AdS}_7\times {\rm S}_4$ case in [13] 
by using the same method that, as we learn from that reference,
was originally employed in [4].

In sect. 1, we present the ${\rm AdS}_7\times {\rm S}_4$
compactification relevant for SCFT$_6$.
Then, in sect. 2, we describe the ${\rm AdS}_4\times {\rm S}_7$
compactification related to SCFT$_3$. 
This case presents a technical complication which 
we avoid by resorting to a dual formulation. 
In sect. 3, we illustrate our conclusions.


\par\vskip.6cm
\item{\bf 1.} {\bf The ${\rm AdS}_7\times {\rm S}_4$ model}
\vskip.4cm
\par
The model considered here is the usual $3$--form formulation of $11d$
supergravity. The basic fields of the bosonic sector
are the metric $g$ and the $3$--form field $A$. 
The bosonic part of the action is 
$$
I={1\over 4\kappa^2}\int_{M_{11}}\Big[R(g)*_g1-F(A)\wedge *_gF(A)
+\hbox{$2^{1\over 2}\over 3$}A\wedge F(A)\wedge F(A)\Big].
\eqno(1.1)
$$
Here, $R(g)$ is the Ricci scalar of the metric $g$.
$F(A)$ is the $4$--form field strength of the $3$--form $A$ 
and is given by
$$
F(A)={\rm d}A.
\eqno(1.2)
$$
The space time $M_{11}$ is taken to have the topology 
${\rm AdS}_7\times {\rm S}_4$.

Then, one can easily verify that the field equations  
admit the standard solution $\bar g_{ij}$,
$\bar A_{ijk}$ generated by the Freund Rubin ansatz [18].
For this, the only non vanishing components of the Riemann tensor 
and field strength are given by
$$
R(\bar g)_{\alpha\beta\gamma\delta}=
-\hbox{$1\over18$}e^2\big(\bar g_{\alpha\gamma}\bar g_{\beta\delta}
-\bar g_{\alpha\delta}\bar g_{\beta\gamma}\big),\quad
R(\bar g)_{\kappa\lambda\mu\nu}=\hbox{$2\over 9$}e^2
\big(\bar g_{\kappa\mu}\bar g_{\lambda\nu}
-\bar g_{\kappa\nu}\bar g_{\lambda\mu}),
\eqno(1.3a)-(1.3b)
$$
$$
F(\bar A)_{\kappa\lambda\mu\nu}=e\bar\epsilon_{\kappa\lambda\mu\nu},
\eqno(1.4)
$$
where $e$ is the compactification scale
\footnote{}{}
\footnote{${}^1$}{In this paper, we adopt the following conventions. 
Latin lower case letters $i,j,k,l,\ldots$ denote $M_{11}$ indices.
Early Greek lower case letters $\alpha,\beta,\gamma,\delta,\ldots$
denote ${\rm AdS}_{7/4}$ indices.
Late Greek lower case letters $\kappa,\lambda,\mu,\nu\dots$
denote ${\rm S}_{4/7}$ indices.}.

We expand the action in fluctuations around the
background $\bar g_{ij}$, $\bar A_{ijk}$.
We parametrize the fluctuations $\delta g_{ij}$, $\delta A_{ijk}$
of the fields $g_{ij}$, $A_{ijk}$ around the background as in [15]
$$
\eqalignno{
\delta g_{\kappa\lambda}&=m_{\kappa\lambda}+\bar\nabla_\kappa n_\lambda
+\bar\nabla_\lambda n_\kappa+(\bar\nabla_\kappa\bar\nabla_\lambda
-\hbox{$1\over 4$}\bar g_{\kappa\lambda}\bar\nabla^\rho\bar\nabla_\rho)p
+\hbox{$1\over 4$}\bar g_{\kappa\lambda}\pi,\vphantom{\Big[}&(1.5a)\cr
m^\rho{}_\rho&=0, \quad \bar\nabla^\rho m_{\rho\kappa}=0,\quad
\bar\nabla^\rho n_\rho=0,\vphantom{\Big[}&\cr
\delta g_{\kappa\alpha}&=k_{\kappa\alpha}+\bar\nabla_\kappa l_\alpha,
\vphantom{\Big[}&(1.5b)\cr
\bar\nabla^\rho k_{\rho\alpha}&=0,\vphantom{\Big[}&\cr
\delta g_{\alpha\beta}&=h_{\alpha\beta}
-\hbox{$1\over 5$}\bar g_{\alpha\beta}\pi,
\vphantom{\Big[}&(1.5c)\cr}
$$
and 
$$
\eqalignno{
\delta A_{\kappa\lambda\mu}&=
3\bar\nabla_{[\kappa}a_{\lambda\mu]}
+\bar\epsilon_{\kappa\lambda\mu}{}^\rho\bar\nabla_\rho b,
\vphantom{\Big[}&(1.6)\cr
\bar\nabla^\rho a_{\rho\kappa}&=0.&\cr}
$$

We find that the quadratic part of the action is given 
by the following expression
$$
\eqalignno{
I_{[2]}&={1\over 4\kappa^2}\int_{{\rm AdS}_7}d^7y (-\bar g_7)^{1\over 2}
 \int_{{\rm S}_4}d^4x (\bar g_4)^{1\over 2}&\cr
&\Big\{\hbox{$1\over 2$}\bar\nabla^\alpha h_{\alpha\gamma}\bar\nabla^\beta
h_\beta{}^\gamma\vphantom{\Big\{}
-\hbox{$1\over 2$}\bar\nabla^\alpha h_{\alpha\beta}\bar\nabla^\beta 
h^\gamma{}_\gamma
+\hbox{$1\over 4$}\bar\nabla^\alpha h^\beta{}_\beta\bar\nabla_\alpha 
h^\gamma{}_\gamma
-\hbox{$1\over 4$}\bar\nabla^\alpha h^{\beta\gamma}\bar\nabla_\alpha 
h_{\beta\gamma}\vphantom{\Big\{}&\cr
&+\hbox{$1\over 4$}\bar\nabla^\kappa h^\beta{}_\beta\bar\nabla_\kappa 
h^\gamma{}_\gamma
-\hbox{$1\over 4$}\bar\nabla^\kappa h^{\beta\gamma}\bar\nabla_\kappa 
h_{\beta\gamma}
+\hbox{$1\over 18$}e^2h^\alpha{}_\alpha h^\beta{}_\beta
+\hbox{$1\over 36$}e^2h^{\alpha\beta}h_{\alpha\beta}\vphantom{\Big\{}&\cr
&-\hbox{$9\over 80$}\bar\nabla^\alpha\pi\bar\nabla_\alpha\pi
-\hbox{$9\over 800$}\bar\nabla^\kappa\pi\bar\nabla_\kappa\pi
-\hbox{$9\over 20$}e^2\pi\pi\vphantom{\Big\{}&\cr
&-\bar\nabla^\alpha\bar\nabla^\kappa b\bar\nabla_\alpha\bar\nabla_\kappa b
-\bar\nabla^\kappa\bar\nabla_\kappa b\bar\nabla^\lambda\bar\nabla_\lambda b
-\hbox{$12\over 5$}e\bar\nabla^\kappa\bar\nabla_\kappa b\pi
\vphantom{\Big\{}&\cr
&+h^\alpha{}_\alpha\big(\hbox{$9\over 40$}\bar\nabla^\kappa\bar\nabla_\kappa\pi
+e\bar\nabla^\kappa\bar\nabla_\kappa b\big)
+\hbox{$1\over 2$}(\bar\nabla_\kappa\bar\nabla_\lambda
-\hbox{$1\over 4$}\bar g_{\kappa\lambda}\bar\nabla^\rho\bar\nabla_\rho)p
\bar\nabla^\kappa\bar\nabla^\lambda\big(h^\alpha{}_\alpha
-\hbox{$9\over 10$}\pi\big)\vphantom{\Big\{}&\cr
&-\bar\nabla^\kappa l^\alpha\bar\nabla_\kappa\big(
\hbox{$9\over 20$}\bar\nabla_\alpha\pi+2e\bar\nabla_\alpha b
+\bar\nabla^\beta h_{\beta\alpha}-\bar\nabla_\alpha h^\beta{}_\beta
\big)+\ldots\Big\}.&(1.7)\cr}
$$
Above, the ellipses denote terms not containing the fields 
$h_{\alpha\beta}$, $\pi$, $b$. Note that the field components 
$m_{\kappa\lambda}$, $n_\kappa$, $k_{\kappa\alpha}$,
$a_{\kappa\lambda}$, $\delta A_{\alpha ij}$
are decoupled at quadratic level from $h_{\alpha\beta}$, $\pi$, $b$. 
Also, $p$ and $l_\alpha$ act as Lagrange multipliers enforcing certain 
constraints. 
We have focused on the fields $\pi, b$ and those mixing with them, 
since they contain the scalar fluctuations which couple 
to the chiral operator of the boundary conformal field theory
of our interest.

At this point, one might partially fix the gauge by imposing
$$
\bar\nabla^\rho\big(\delta g_{\rho\kappa}
-\hbox{$1\over 4$}\bar g_{\rho\kappa}\delta g^\sigma{}_\sigma\big)=0,\quad
\bar\nabla^\rho\delta g_{\rho\alpha}=0;
\eqno(1.8a)-(1.8b)
$$
$$
\bar\nabla^\rho\delta A_{ij\rho}=0,
\eqno(1.9)
$$
as shown in [15]. These conditions imply in particular that
$p=0$ and $l_\alpha=0$ and so, upon gauge fixing, 
the last two terms of $I_{[2]}$ vanish. 
The corresponding constraints must then be enforced by hand.

However, instead of fixing the gauge at this stage, we prefer first
to isolate the relevant scalar degrees of freedom by performing
field redefinitions, as done in [14]. We write the AdS metric 
fluctuations as
$$
h_{\alpha\beta}=\phi_{\alpha\beta}
+\eta  \bar g_{\alpha\beta} 
+ \bar\nabla_\alpha\bar\nabla_\beta \zeta 
\eqno(1.10)
$$
with $\eta$ and $\zeta$ chosen in such a way to decouple at the quadratic 
level the field $\phi_{\alpha\beta}$ from the $\pi$--$b$ sector.
We find 
$$
h_{\alpha\beta}=\phi_{\alpha\beta}
+\big(\bar\nabla_\alpha\bar\nabla_\beta
-\hbox{$1\over 5$}\bar g_{\alpha\beta}
\bar\nabla^\kappa\bar\nabla_\kappa\big)\varphi,
\eqno(1.11a)
$$
where
$$
\varphi=
\big(-\bar\nabla^\lambda\bar\nabla_\lambda+\hbox{$5\over 18$}e^2\big)^{-1}
\big(\hbox{$3\over 8$}\pi+\hbox{$5\over 3$}eb\big).
\eqno(1.11b)
$$
The action $I_{[2]}$ then becomes
$$
\eqalignno{
I_{[2]}&={1\over 4\kappa^2}\int_{{\rm AdS}_7}d^7y (-\bar g_7)^{1\over 2}
 \int_{{\rm S}_4}d^4x (\bar g_4)^{1\over 2}&\cr
&\Big\{\hbox{$1\over 2$}\bar\nabla^\alpha \phi_{\alpha\gamma}\bar\nabla^\beta
\phi_\beta{}^\gamma\vphantom{\Big\{}
-\hbox{$1\over 2$}\bar\nabla^\alpha \phi_{\alpha\beta}\bar\nabla^\beta 
\phi^\gamma{}_\gamma
+\hbox{$1\over 4$}\bar\nabla^\alpha \phi^\beta{}_\beta\bar\nabla_\alpha 
\phi^\gamma{}_\gamma
-\hbox{$1\over 4$}\bar\nabla^\alpha \phi^{\beta\gamma}\bar\nabla_\alpha 
\phi_{\beta\gamma}\vphantom{\Big\{}&\cr
&+\hbox{$1\over 4$}\bar\nabla^\kappa \phi^\beta{}_\beta\bar\nabla_\kappa 
\phi^\gamma{}_\gamma
-\hbox{$1\over 4$}\bar\nabla^\kappa \phi^{\beta\gamma}\bar\nabla_\kappa 
\phi_{\beta\gamma}
+\hbox{$1\over 18$}e^2\phi^\alpha{}_\alpha \phi^\beta{}_\beta
+\hbox{$1\over 36$}e^2\phi^{\alpha\beta}\phi_{\alpha\beta}
\vphantom{\Big\{}&\cr
&-\hbox{$3\over 10$}\bar\nabla^\kappa\bar\nabla_\kappa
\big(-\bar\nabla^\lambda\bar\nabla_\lambda+\hbox{$5\over 18$}e^2\big)^{-1}
\bar\nabla^\alpha\big(\hbox{$3\over 8$}\pi+\hbox{$5\over 3$}eb\big)
\bar\nabla_\alpha\big(\hbox{$3\over 8$}\pi+\hbox{$5\over 3$}eb\big)
\vphantom{\Big\{}&\cr
&-\hbox{$21\over 50$}\bar\nabla^\kappa\bar\nabla_\kappa
\bar\nabla^\lambda\bar\nabla_\lambda
\big(-\bar\nabla^\mu\bar\nabla_\mu+\hbox{$5\over 18$}e^2\big)^{-1}
\big(\hbox{$3\over 8$}\pi+\hbox{$5\over 3$}eb\big)
\big(\hbox{$3\over 8$}\pi+\hbox{$5\over 3$}eb\big)
\vphantom{\Big\{}&\cr
&-\hbox{$9\over 80$}\bar\nabla^\alpha\pi\bar\nabla_\alpha\pi
-\hbox{$9\over 800$}\bar\nabla^\kappa\pi\bar\nabla_\kappa\pi
-\hbox{$9\over 20$}e^2\pi\pi\vphantom{\Big\{}&\cr
&-\bar\nabla^\alpha\bar\nabla^\kappa b\bar\nabla_\alpha\bar\nabla_\kappa b
-\bar\nabla^\kappa\bar\nabla_\kappa b\bar\nabla^\lambda\bar\nabla_\lambda b
-\hbox{$12\over 5$}e\bar\nabla^\kappa\bar\nabla_\kappa b\pi
\vphantom{\Big\{}&\cr
&+\hbox{$1\over 2$}(\bar\nabla_\kappa\bar\nabla_\lambda
-\hbox{$1\over 4$}\bar g_{\kappa\lambda}\bar\nabla^\rho\bar\nabla_\rho)p
\bar\nabla^\kappa\bar\nabla^\lambda\big[\phi^\alpha{}_\alpha
-\hbox{$9\over 10$}\pi\vphantom{\Big\{}&\cr
&+\big(\bar\nabla^\alpha\bar\nabla_\alpha
-\hbox{$7\over 5$}\bar\nabla^\kappa\bar\nabla_\kappa\big)
\big(-\bar\nabla^\lambda\bar\nabla_\lambda+\hbox{$5\over 18$}e^2\big)^{-1}
\big(\hbox{$3\over 8$}\pi+\hbox{$5\over 3$}eb\big)\big]\vphantom{\Big\{}&\cr
&-\bar\nabla^\kappa l^\alpha\bar\nabla_\kappa\big(
\bar\nabla^\beta\phi_{\beta\alpha}-\bar\nabla_\alpha\phi^\beta{}_\beta
\big)+\ldots\Big\}.&(1.12)\cr}
$$

Let us concentrate on the scalar $\pi$--$b$ sector of the above action.
We expand the fields $\pi$, $b$ in harmonics $Y^{(4)}_I$ of the ${\rm S}_4$ 
d'Alembertian $\bar\nabla^\kappa\bar\nabla_\kappa$
$$
\bar\nabla^\kappa\bar\nabla_\kappa Y^{(4)}_I=
-\hbox{$2\over 9$}e^2k(k+3)Y^{(4)}_I,
\eqno(1.13)
$$
where $k=0,1,2,\ldots$ depends on $I$. The $Y^{(4)}_I$ are 
conventionally normalized as
$$
\eqalignno{
\int_{{\rm S}_4}(g_4)^{1\over 2}Y^{(4)}_IY^{(4)}_J
&=(\hbox{$2\over 9$}e^2)^{-2}z^{(4)}(k)\delta_{IJ},&(1.14a)\cr
z^{(4)}(k)&={8\pi^2k!\over (2k+3)!}.&(1.14b)\cr}
$$
The expansions read
$$
\phi_{\alpha\beta}=\sum_I\phi_{I\alpha\beta}Y^{(4)}_I,\quad 
\pi=\sum_I\pi_IY^{(4)}_I,\quad b=\sum_I b_IY^{(4)}_I,
$$
$$
p=\sum_I p_IY^{(4)}_I,\quad
l_\alpha=\sum_I l_{I\alpha}Y^{(4)}_I.
\eqno(1.15a)-(1.15e)
$$
Now, to diagonalize the kinetic terms,
one introduces new scalars $s_I$, $t_I$ such that
$$
\eqalignno{
\pi_I&=\hbox{$4\over 3$}ks_I+\hbox{$4\over 3$}(k+3)t_I,&(1.16a)\cr
b_I&=\hbox{$3\over 4$}e^{-1}\big(s_I-t_I\big).&(1.16b)\cr}
$$
In terms of $s_I$, $t_I$, $I_{[2]}$ takes the simple form
$$
\eqalignno{
I_{[2]}&={1\over 4\kappa^2}(\hbox{$2\over 9$}e^2)^{-2}
\int_{{\rm AdS}_7}d^7y (-\bar g_7)^{1\over 2}\sum_I z^{(4)}(k)&\cr
&\Big\{\hbox{$1\over 2$}\bar\nabla^\alpha \phi_{I\alpha\gamma}\bar\nabla^\beta
\phi_{I\beta}{}^\gamma\vphantom{\Big\{}
-\hbox{$1\over 2$}\bar\nabla^\alpha \phi_{I\alpha\beta}\bar\nabla^\beta 
\phi_I{}^\gamma{}_\gamma
+\hbox{$1\over 4$}\bar\nabla^\alpha \phi_I{}^\beta{}_\beta\bar\nabla_\alpha 
\phi_I{}^\gamma{}_\gamma
-\hbox{$1\over 4$}\bar\nabla^\alpha \phi_I{}^{\beta\gamma}\bar\nabla_\alpha 
\phi_{I\beta\gamma}\vphantom{\Big\{}&\cr
&+\hbox{$1\over 4$}m_{\phi I}'{}^2
\phi_I{}^\alpha{}_\alpha \phi_I{}^\beta{}_\beta
-\hbox{$1\over 4$}m_{\phi I}''{}^2\phi_I{}^{\alpha\beta}\phi_{I\alpha\beta}
\vphantom{\Big\{}&\cr
&+A_{sI}\Big[-\hbox{$1\over 2$}\bar\nabla^\alpha s_I\bar\nabla_\alpha s_I
-\hbox{$1\over 2$}m_{sI}{}^2s_Is_I\Big]\vphantom{\Big\{}
+A_{tI}\Big[-\hbox{$1\over 2$}\bar\nabla^\alpha t_I\bar\nabla_\alpha t_I
-\hbox{$1\over 2$}m_{tI}{}^2t_It_I\Big]
\vphantom{\Big\{}&\cr
&+u_Ip_I\Big[\phi_I{}^\alpha{}_\alpha
+v_{sI}\Big(\bar\nabla^\alpha\bar\nabla_\alpha s_I-m_{sI}{}^2s_I\Big)
+v_{tI}\Big(\bar\nabla^\alpha\bar\nabla_\alpha t_I-m_{tI}{}^2t_I\Big)\Big]
\vphantom{\Big\{}&\cr
&-w_Il_I{}^\alpha\Big[\bar\nabla^\beta\phi_{I\beta\alpha}
-\bar\nabla_\alpha\phi_I{}^\beta{}_\beta\Big]+\ldots\Big\},
&(1.17)\cr}
$$
where
$$\eqalignno{
A_{sI}&={(2k+3)k(k-1)\over 2(2k+1)},&(1.18a)\cr
A_{tI}&={(2k+3)(k+3)(k+4)\over 2(2k+5)},&(1.18b)\cr
m_{\phi I}'{}^2&=\hbox{$2\over 9$}e^2(k^2+3k+1),\vphantom{9\over 8}&(1.19a)\cr
m_{\phi I}''{}^2&=\hbox{$2\over 9$}e^2(k^2+3k-\hbox{$1\over 2$}),
\vphantom{9\over 8}&(1.19b)\cr
m_{sI}{}^2&=\hbox{$2\over 9$}e^2k(k-3),\vphantom{9\over 8}&(1.20a)\cr
m_{tI}{}^2&=\hbox{$2\over 9$}e^2(k+3)(k+6),\vphantom{9\over 8}&(1.20b)\cr
u_I&=(\hbox{$2\over 9$}e^2)^2\hbox{$3\over 8$}(k-1)k(k+3)(k+4),
\vphantom{9\over 8}&(1.21)\cr
v_{sI}&=(\hbox{$2\over 9$}e^2)^{-1}{1\over (2k+1)},&(1.22a)\cr
v_{tI}&=(\hbox{$2\over 9$}e^2)^{-1}{1\over (2k+5)},&(1.22b)\cr
w_I&=\hbox{$2\over 9$}e^2k(k+3).&(1.23)\cr}
$$
Note that the modes $s_I$ with $k=0,1$ do not appear, so these are 
gauge degrees of freedom. 

Fixing the gauge involves setting $p_I=0$, $l_{I\alpha}=0$. Thus, after gauge 
fixing, $s_I$, $t_I$ are free fields. On the $s_I$, $t_I$ mass shell
the constraints are simply
$$\eqalignno{
\phi_I{}^\alpha{}_\alpha&=0,\quad k\geq 2,&(1.24a)\cr
&&\cr
\bar\nabla^\beta\phi_{I\beta\alpha}&=0,\quad k\geq 1.&(1.24b)\cr}
$$

As a check of our results we note that we have reproduced the masses 
worked out in [15]. The particular
coefficient $A_{sI}$ has also been deduced recently in [13]
by using a consistency requirement involving other normalizations and
the values of certain three point couplings.
We confirm that result as well.


\par\vskip.6cm
\item{\bf 2.} {\bf The ${\rm AdS}_4\times {\rm S}_7$ model}
\vskip.4cm
\par
We now analyze the 11d supergravity compactification on 
${\rm AdS}_4\times {\rm S}_7$. The relevant action is again (1.1), where 
now $M_{11}$ has the topology ${\rm AdS}_4\times {\rm S}_7$.
The field equations admit the standard solution $\bar g_{ij}$,
$\bar A_{ijk}$ generated by the Freund Rubin type ansatz [18].
On this background, the only non vanishing components of the Riemann tensor 
and field strength are given by
$$
R(\bar g)_{\alpha\beta\gamma\delta}=
-\hbox{$2\over 9$}e^2\big(\bar g_{\alpha\gamma}\bar g_{\beta\delta}
-\bar g_{\alpha\delta}\bar g_{\beta\gamma}\big),\quad
R(\bar g)_{\kappa\lambda\mu\nu}=\hbox{$1\over 18$}e^2
\big(\bar g_{\kappa\mu}\bar g_{\lambda\nu}
-\bar g_{\kappa\nu}\bar g_{\lambda\mu}),
\eqno(2.1a)-(2.1b)
$$
$$
F(\bar A)_{\alpha\beta\gamma\delta}
=-e\bar\epsilon_{\alpha\beta\gamma\delta},
\eqno(2.2)
$$
where $e$ is the compactification scale.

We parametrize the metric fluctuations $\delta g_{ij}$ around the background 
as follows [17]
$$
\eqalignno{
\delta g_{\kappa\lambda}&=m_{\kappa\lambda}+\bar\nabla_\kappa n_\lambda
+\bar\nabla_\lambda n_\kappa+(\bar\nabla_\kappa\bar\nabla_\lambda
-\hbox{$1\over 7$}\bar g_{\kappa\lambda}\bar\nabla^\rho\bar\nabla_\rho)p
+\hbox{$1\over 7$}\bar g_{\kappa\lambda}\pi,\vphantom{\Big[}&(2.3a)\cr
m^\rho{}_\rho&=0, \quad \bar\nabla^\rho m_{\rho\kappa}=0,\quad
\bar\nabla^\rho n_\rho=0,\vphantom{\Big[}&\cr
\delta g_{\kappa\alpha}&=k_{\kappa\alpha}+\bar\nabla_\kappa l_\alpha,
\vphantom{\Big[}&(2.3b)\cr
\bar\nabla^\rho k_{\rho\alpha}&=0,\vphantom{\Big[}&\cr
\delta g_{\alpha\beta}&=h_{\alpha\beta}
-\hbox{$1\over 2$}\bar g_{\alpha\beta}\pi.
\vphantom{\Big[}&(2.3c)\cr}
$$
As to the 3-form fluctuations, it is sufficient for our purposes
to use 
$$
b^\alpha=-\hbox{$1\over 3!$}\bar\epsilon^{\alpha\beta\gamma\delta}
\delta A_{\beta\gamma\delta}
\eqno(2.4)
$$
and denote by $a_{\kappa\gamma\delta}$ the longitudinal (in ${\rm S}_7$)
components of $\delta A_{\kappa\gamma\delta}$.
Then, the part of the action 
quadratic in fluctuations is given by the following expression
$$
\eqalignno{
I_{[2]}&={1\over 4\kappa^2}\int_{{\rm AdS}_4}d^4y (-\bar g_4)^{1\over 2}
 \int_{{\rm S}_7}d^7x (\bar g_7)^{1\over 2}&\cr
&\Big\{\hbox{$1\over 2$}\bar\nabla^\alpha h_{\alpha\gamma}\bar\nabla^\beta
h_\beta{}^\gamma\vphantom{\Big\{}
-\hbox{$1\over 2$}\bar\nabla^\alpha h_{\alpha\beta}\bar\nabla^\beta 
h^\gamma{}_\gamma
+\hbox{$1\over 4$}\bar\nabla^\alpha h^\beta{}_\beta\bar\nabla_\alpha 
h^\gamma{}_\gamma
-\hbox{$1\over 4$}\bar\nabla^\alpha h^{\beta\gamma}\bar\nabla_\alpha 
h_{\beta\gamma}\vphantom{\Big\{}&\cr
&+\hbox{$1\over 4$}\bar\nabla^\kappa h^\beta{}_\beta\bar\nabla_\kappa 
h^\gamma{}_\gamma
-\hbox{$1\over 4$}\bar\nabla^\kappa h^{\beta\gamma}\bar\nabla_\kappa 
h_{\beta\gamma}
+\hbox{$11\over 36$}e^2h^\alpha{}_\alpha h^\beta{}_\beta
+\hbox{$1\over 9$}e^2h^{\alpha\beta}h_{\alpha\beta}\vphantom{\Big\{}&\cr
&-\hbox{$9\over 56$}\bar\nabla^\alpha\pi\bar\nabla_\alpha\pi
+\hbox{$9\over 196$}\bar\nabla^\kappa\pi\bar\nabla_\kappa\pi
+\hbox{$45\over 28$}e^2\pi\pi\vphantom{\Big\{}&\cr
&+\bar\nabla^\alpha b_\alpha\bar\nabla^\beta b_\beta
+\bar\nabla^\kappa b^\alpha \bar\nabla_\kappa b_\alpha
-3e\bar\nabla^\alpha b_\alpha\pi
\vphantom{\Big\{}&\cr
&+h^\alpha{}_\alpha\big(\hbox{$9\over 28$}\bar\nabla^\kappa\bar\nabla_\kappa\pi
-\hbox{$3\over 2$}e^2\pi+e\bar\nabla^\beta b_\beta\big)
+\hbox{$1\over 2$}(\bar\nabla_\kappa\bar\nabla_\lambda
-\hbox{$1\over 7$}\bar g_{\kappa\lambda}\bar\nabla^\rho\bar\nabla_\rho)p
\bar\nabla^\kappa\bar\nabla^\lambda\big(h^\alpha{}_\alpha
-\hbox{$9\over 7$}\pi\big)\vphantom{\Big\{}&\cr
&-\bar\nabla^\kappa l^\alpha\bar\nabla_\kappa\big(
\hbox{$9\over 14$}\bar\nabla_\alpha\pi-2eb_\alpha
+\bar\nabla^\beta h_{\beta\alpha}-\bar\nabla_\alpha h^\beta{}_\beta
\big)+\bar\nabla^\kappa a_{\kappa\gamma\delta}
\bar\epsilon^{\alpha\beta\gamma\delta}\bar\nabla_\alpha b_\beta
+\ldots\Big\},&(2.5)\cr}
$$
where  the dots indicate a totally decoupled sector which is not relevant
for our purposes.
In fact, the scalar fields of interest are contained only in $\pi$
and $b_\alpha$ and those fields mixing with them. One can check this
statement by analyzing the corresponding field equations  
$$
\eqalignno{
&\hbox{$9\over 28$}\bar\nabla^\alpha\bar\nabla_\alpha\pi
-\hbox{$9\over 98$}\bar\nabla^\kappa\bar\nabla_\kappa\pi
+\hbox{$45\over 14$}e^2\pi\pi-3e\bar\nabla^\alpha b_\alpha
+\big(\hbox{$9\over 28$}\bar\nabla^\kappa\bar\nabla_\kappa\pi
-\hbox{$3\over 2$}e^2\pi\big)h^\alpha{}_\alpha=0, 
\vphantom{\Big[}&(2.6a)\cr
&2\bar\nabla^\alpha\bar\nabla^\beta b_\beta
+2\bar\nabla^\kappa\bar\nabla_\kappa b^\alpha
-3e\bar\nabla^\alpha\pi+e\bar\nabla^\alpha h^\beta{}_\beta=0, 
\vphantom{\Big[}&(2.6b)\cr
&h^\alpha{}_\alpha-\hbox{$9\over 7$}\pi=0, 
\vphantom{\Big[}&(2.6c)\cr
&\bar\nabla^\alpha b^\beta-\bar\nabla^\beta b^\alpha=0. 
\vphantom{\Big[}&(2.6d)\cr}
$$
The last two equations are constraints arising from imposing gauge
fixing conditions analogous to (1.8a)--(1.8b) and (1.9), 
which imply in particular $p=0$, $a_{\kappa\gamma\delta}=0$ 
\footnote{}{}
\footnote{${}^2$}{
To be precise, (2.6c)--(2.6d) holds only 
up to zero modes of the operators $\bar\nabla_\kappa\bar\nabla_\lambda
-\hbox{$1\over 7$}\bar g_{\kappa\lambda}\bar\nabla^\rho\bar\nabla_\rho$ and 
$\bar\nabla_\kappa$, respectively [15]. 
We will neglect this technical complications in the following discussion 
for the sake of simplicity.}. Substituting the solution of 
the constraints in the first two equations, one gets
$$
\eqalignno{
&\bar\nabla^\alpha\bar\nabla_\alpha\pi
-\hbox{$28\over 3$}e\bar\nabla^\alpha\bar\nabla_\alpha b
+\bar\nabla^\kappa\bar\nabla_\kappa\pi+4e^2\pi=0, 
\vphantom{\Big[}&(2.7a)\cr
&\bar\nabla^\alpha\bar\nabla_\alpha b+\bar\nabla^\kappa\bar\nabla_\kappa b
-\hbox{$6\over 7$}e\pi=0, 
\vphantom{\Big[}&(2.7b)\cr}
$$
where $b$ is defined by the relation $b^\alpha=\bar\nabla^\alpha b$,
which solves (2.6d). Thus, the above action describes a symmetric rank 2 
tensor $h_{\alpha\beta}$ and two scalar fields $\pi$, $b$.
One can expand them in spherical harmonics and diagonalize the field 
equations (2.7a)--(2.7b) to obtain the massive Kaluza--Klein towers of 
scalars of interest [17].

However, we now face a technical difficulty. It is apparently hard to 
diagonalize the action (2.5) by performing field redefinitions.
The reason for this is that, as shown above, the vector field 
$b_\alpha$ describes simply a scalar field $b$. A field redefinition replacing 
the vector field by the scalar one is a form of duality transformation.
Typically, such transformations exchange field equations and Bianchi
identities and are not easily done at the level of action in a consistent 
way by simply substituting in some of the field equations. 
In our case, however, this goal can be achieved by rewriting
the original 11d supergravity action introducing appropriate
auxiliary fields. These, when eliminated in a judicious way, 
realize the desired duality transformation.

Explicitly, instead of the customary $3$--form formulation of 
11d supergravity, we shall adopt the dual $3/6$--form formulation 
worked out in [19--20]. In this approach, the basic fields of the bosonic 
sector are the metric $g$, the $3$--form field $A$ and the $6$--form 
field $C$. The bosonic part of the action now reads 
$$
I={1\over 4\kappa^2}\int_{M_{11}}\Big[R(g)*_g1-H(A,C)\wedge *_gH(A,C)
-\hbox{$2^{1\over 2}2\over 3$}A\wedge F(A)\wedge F(A)\Big].
\eqno(2.8)
$$
$R(g)$ is the Ricci scalar of the metric $g$.
$F(A)$, $H(A,C)$ are the $4$--form field strength of the $3$--form $A$ 
and the $7$--form field strength of the $6$--form $C$
and are given by
$$
F(A)={\rm d}A, \quad H(A,C)={\rm d}C+\hbox{$1\over 2^{1/2}$}A\wedge{\rm d}A.
\eqno(2.9)
$$
The model must be supplemented with the duality constraint
$$
*_g H(A,C)+F(A)=0.
\eqno(2.10)
$$
Using additional auxiliary/gauge fields, one can incorporate this
constraint directly at the level of the action [20], but this
will not be necessary for our purposes. 

One can easily verify that the field equations admit a standard solution
$\bar g_{ij}$, $\bar A_{ijk}$, $\bar C_{ijklmn}$ generated by a Freund Rubin 
type ansatz [18]. For this, the only non vanishing components of the Riemann 
tensor and the $4$--from field strength are as in (2.1a)--(2.1b), (2.2), 
while the $7$--form field strength is given by
$$
H(\bar A,\bar C)_{\kappa\lambda\mu\nu\rho\sigma\tau}
=e\bar\epsilon_{\kappa\lambda\mu\nu\rho\sigma\tau}.
\eqno(2.11)
$$

We expand the action (2.8) in fluctuations around the above background. 
We parame\-tri\-ze the fluctuations $\delta g_{ij}$ as in (2.3a)--(2.3c)
and write the relevant fluctuations $\delta C_{ijklmn}$ as
$$
\eqalignno{
\delta C_{\kappa\lambda\mu\nu\rho\sigma}&=
6\bar\nabla_{[\kappa}a_{\lambda\mu\nu\rho\sigma]}
+\bar\epsilon_{\kappa\lambda\mu\nu\rho\sigma}{}^\tau\bar\nabla_\tau b,
\vphantom{\Big[}&(2.12)\cr
\bar\nabla^\rho a_{\rho\kappa\lambda\mu\nu}&=0.&\cr}
$$

The fluctuations $\delta A_{ijk}$ are not independent from  
$\delta g_{ij}$, $\delta C_{ijklmn}$ because of the constraint 
(2.10). We have explicitly checked, however, that the components  
of $\delta A_{ijk}$ that couple in the action and in the constraint
to the fields $h_{\alpha\beta}$, $\pi$, $b$ give a vanishing contribution 
to the quadratic action of these latter fields, once the constraint is
taken into account. 

Therefore, the quadratic action of the fields $h_{\alpha\beta}$, $\pi$, $b$
could be evaluated by setting formally $\delta A_{ijk}=0$ from the beginning. 
The expression we obtain is 
$$
\eqalignno{
I_{[2]}&={1\over 4\kappa^2}\int_{{\rm AdS}_4}d^4y (-\bar g_4)^{1\over 2}
 \int_{{\rm S}_7}d^7x (\bar g_7)^{1\over 2}&\cr
&\Big\{\hbox{$1\over 2$}\bar\nabla^\alpha h_{\alpha\gamma}\bar\nabla^\beta
h_\beta{}^\gamma\vphantom{\Big\{}
-\hbox{$1\over 2$}\bar\nabla^\alpha h_{\alpha\beta}\bar\nabla^\beta 
h^\gamma{}_\gamma
+\hbox{$1\over 4$}\bar\nabla^\alpha h^\beta{}_\beta\bar\nabla_\alpha 
h^\gamma{}_\gamma
-\hbox{$1\over 4$}\bar\nabla^\alpha h^{\beta\gamma}\bar\nabla_\alpha 
h_{\beta\gamma}\vphantom{\Big\{}&\cr
&+\hbox{$1\over 4$}\bar\nabla^\kappa h^\beta{}_\beta\bar\nabla_\kappa 
h^\gamma{}_\gamma
-\hbox{$1\over 4$}\bar\nabla^\kappa h^{\beta\gamma}\bar\nabla_\kappa 
h_{\beta\gamma}
+\hbox{$1\over 18$}e^2h^\alpha{}_\alpha h^\beta{}_\beta
+\hbox{$1\over 9$}e^2h^{\alpha\beta}h_{\alpha\beta}\vphantom{\Big\{}&\cr
&-\hbox{$9\over 56$}\bar\nabla^\alpha\pi\bar\nabla_\alpha\pi
+\hbox{$9\over 196$}\bar\nabla^\kappa\pi\bar\nabla_\kappa\pi
-\hbox{$9\over 14$}e^2\pi\pi\vphantom{\Big\{}&\cr
&-\bar\nabla^\alpha\bar\nabla^\kappa b\bar\nabla_\alpha\bar\nabla_\kappa b
-\bar\nabla^\kappa\bar\nabla_\kappa b\bar\nabla^\lambda\bar\nabla_\lambda b
+3e\bar\nabla^\kappa\bar\nabla_\kappa b\pi
\vphantom{\Big\{}&\cr
&+h^\alpha{}_\alpha\big(\hbox{$9\over 28$}\bar\nabla^\kappa\bar\nabla_\kappa\pi
-e\bar\nabla^\kappa\bar\nabla_\kappa b\big)
+\hbox{$1\over 2$}(\bar\nabla_\kappa\bar\nabla_\lambda
-\hbox{$1\over 7$}\bar g_{\kappa\lambda}\bar\nabla^\rho\bar\nabla_\rho)p
\bar\nabla^\kappa\bar\nabla^\lambda\big(h^\alpha{}_\alpha
-\hbox{$9\over 7$}\pi\big)\vphantom{\Big\{}&\cr
&-\bar\nabla^\kappa l^\alpha\bar\nabla_\kappa\big(
\hbox{$9\over 14$}\bar\nabla_\alpha\pi-2e\bar\nabla_\alpha b
+\bar\nabla^\beta h_{\beta\alpha}-\bar\nabla_\alpha h^\beta{}_\beta
\big)+\ldots\Big\}.&(2.13)\cr}
$$
Above, the ellipses denote terms not containing the fields 
$h_{\alpha\beta}$, $\pi$, $b$. Note that the field components 
$m_{\kappa\lambda}$, $n_\kappa$, $k_{\kappa\alpha}$,
$a_{\kappa\lambda\mu\nu\rho}$, $\delta C_{\alpha ijklm}$
are decoupled at quadratic level from $h_{\alpha\beta}$, $\pi$, $b$. 
Also, $p$ and $l_\alpha$ act as Lagrange multipliers enforcing certain 
constraints. 

The gauge is partially fixed by imposing
$$
\bar\nabla^\rho\big(\delta g_{\rho\kappa}
-\hbox{$1\over 7$}\bar g_{\rho\kappa}\delta g^\sigma{}_\sigma\big)=0,\quad
\bar\nabla^\rho\delta g_{\rho\alpha}=0;
\eqno(2.14a)-(2.14b)
$$
$$
\bar\nabla^\rho\delta C_{ijklm\rho}=0,
\eqno(2.15)
$$
as in [17]. These imply in particular that
$p=0$, $l_\alpha=0$. 
Fixing the gauge as indicated, the last two terms of $I_{[2]}$ vanish. 
The corresponding constraints must then be enforced by hand.

However, instead of fixing the gauge at this stage, we prefer first
to isolate the relevant scalar degrees of freedom by performing
field redefinitions, as done in the previous section.
Following again [14], we perform a field redefinition of the form (1.12)
with $\eta$ and $\zeta$ chosen in such a way to decouple at the quadratic 
level the field $\phi_{\alpha\beta}$ from the $\pi$--$b$ sector. One finds 
in this way that
$$
h_{\alpha\beta}=\phi_{\alpha\beta}
+\big(\bar\nabla_\alpha\bar\nabla_\beta
-\hbox{$1\over 2$}\bar g_{\alpha\beta}
\bar\nabla^\kappa\bar\nabla_\kappa\big)\varphi,
\eqno(2.16a)
$$
where
$$
\varphi=
\big(-\bar\nabla^\lambda\bar\nabla_\lambda+\hbox{$4\over 9$}e^2\big)^{-1}
\big(\hbox{$3\over 7$}\pi-\hbox{$4\over 3$}eb\big).
\eqno(2.16b)
$$
The action $I_{[2]}$ then becomes
$$
\eqalignno{
I_{[2]}&={1\over 4\kappa^2}\int_{{\rm AdS}_4}d^4y (-\bar g_4)^{1\over 2}
 \int_{{\rm S}_7}d^7x (\bar g_7)^{1\over 2}&\cr
&\Big\{\hbox{$1\over 2$}\bar\nabla^\alpha \phi_{\alpha\gamma}\bar\nabla^\beta
\phi_\beta{}^\gamma\vphantom{\Big\{}
-\hbox{$1\over 2$}\bar\nabla^\alpha \phi_{\alpha\beta}\bar\nabla^\beta 
\phi^\gamma{}_\gamma
+\hbox{$1\over 4$}\bar\nabla^\alpha \phi^\beta{}_\beta\bar\nabla_\alpha 
\phi^\gamma{}_\gamma
-\hbox{$1\over 4$}\bar\nabla^\alpha \phi^{\beta\gamma}\bar\nabla_\alpha 
\phi_{\beta\gamma}\vphantom{\Big\{}&\cr
&+\hbox{$1\over 4$}\bar\nabla^\kappa \phi^\beta{}_\beta\bar\nabla_\kappa 
\phi^\gamma{}_\gamma
-\hbox{$1\over 4$}\bar\nabla^\kappa \phi^{\beta\gamma}\bar\nabla_\kappa 
\phi_{\beta\gamma}
+\hbox{$1\over 18$}e^2\phi^\alpha{}_\alpha \phi^\beta{}_\beta
+\hbox{$1\over 9$}e^2\phi^{\alpha\beta}\phi_{\alpha\beta}
\vphantom{\Big\{}&\cr
&-\hbox{$3\over 8$}\bar\nabla^\kappa\bar\nabla_\kappa
\big(-\bar\nabla^\lambda\bar\nabla_\lambda+\hbox{$4\over 9$}e^2\big)^{-1}
\bar\nabla^\alpha\big(\hbox{$3\over 7$}\pi-\hbox{$4\over 3$}eb\big)
\bar\nabla_\alpha\big(\hbox{$3\over 7$}\pi-\hbox{$4\over 3$}eb\big)
\vphantom{\Big\{}&\cr
&-\hbox{$3\over 4$}\bar\nabla^\kappa\bar\nabla_\kappa
\bar\nabla^\lambda\bar\nabla_\lambda
\big(-\bar\nabla^\mu\bar\nabla_\mu+\hbox{$4\over 9$}e^2\big)^{-1}
\big(\hbox{$3\over 7$}\pi-\hbox{$4\over 3$}eb\big)
\big(\hbox{$3\over 7$}\pi-\hbox{$4\over 3$}eb\big)
\vphantom{\Big\{}&\cr
&-\hbox{$9\over 56$}\bar\nabla^\alpha\pi\bar\nabla_\alpha\pi
+\hbox{$9\over 196$}\bar\nabla^\kappa\pi\bar\nabla_\kappa\pi
-\hbox{$9\over 14$}e^2\pi\pi\vphantom{\Big\{}&\cr
&-\bar\nabla^\alpha\bar\nabla^\kappa b\bar\nabla_\alpha\bar\nabla_\kappa b
-\bar\nabla^\kappa\bar\nabla_\kappa b\bar\nabla^\lambda\bar\nabla_\lambda b
+3e\bar\nabla^\kappa\bar\nabla_\kappa b\pi
\vphantom{\Big\{}&\cr
&+\hbox{$1\over 2$}(\bar\nabla_\kappa\bar\nabla_\lambda
-\hbox{$1\over 7
$}\bar g_{\kappa\lambda}\bar\nabla^\rho\bar\nabla_\rho)p
\bar\nabla^\kappa\bar\nabla^\lambda\big[\phi^\alpha{}_\alpha
-\hbox{$9\over 7$}\pi\vphantom{\Big\{}&\cr
&+\big(\bar\nabla^\alpha\bar\nabla_\alpha
-2\bar\nabla^\kappa\bar\nabla_\kappa\big)
\big(-\bar\nabla^\lambda\bar\nabla_\lambda+\hbox{$4\over 9$}e^2\big)^{-1}
\big(\hbox{$3\over 7$}\pi-\hbox{$4\over 3$}eb\big)\big]\vphantom{\Big\{}&\cr
&-\bar\nabla^\kappa l^\alpha\bar\nabla_\kappa\big(
\bar\nabla^\beta\phi_{\beta\alpha}-\bar\nabla_\alpha\phi^\beta{}_\beta
\big)+\ldots\Big\}.&(2.17)\cr}
$$

Let us concentrate on the scalar $\pi$--$b$ sector of the above action.
Proceeding as in the previous section, 
we expand the fields $\pi$, $b$ in harmonics $Y^{(7)}_I$ of the ${\rm S}_7$ 
d'Alembertian $\bar\nabla^\kappa\bar\nabla_\kappa$
$$
\bar\nabla^\kappa\bar\nabla_\kappa Y^{(7)}_I=
-\hbox{$1\over 18$}e^2k(k+6)Y^{(7)}_I,
\eqno(2.18)
$$
where $k=0,1,2,\ldots$ depends on $I$. The $Y^{(7)}_I$ are 
conventionally normalized as
$$
\eqalignno{
\int_{{\rm S}_7}(g_7)^{1\over 2}Y^{(7)}_IY^{(7)}_J
&=(\hbox{$1\over 18$}e^2)^{-{7\over 2}}z^{(7)}(k)\delta_{IJ},&(2.19a)\cr
z^{(7)}(k)&={\pi^4\over 2^{k-1}(k+1)(k+2)(k+3)}.&(2.19b)\cr}
$$
The expansions read
$$
\phi_{\alpha\beta}=\sum_I\phi_{I\alpha\beta}Y^{(7)}_I,\quad 
\pi=\sum_I\pi_IY^{(7)}_I,\quad  
b=\sum_I b_IY^{(7)}_I,
$$
$$
p=\sum_I p_IY^{(7)}_I, \quad 
l_\alpha=\sum_I l_{I\alpha}Y^{(7)}_I.
\eqno(2.20a)-(2.20e)
$$
One introduces new scalars $s_I$, $t_I$ such that
$$
\eqalignno{
\pi_I&=\hbox{$7\over 3$}ks_I+\hbox{$7\over 3$}(k+6)t_I,&(2.21a)\cr
b_I&=3e^{-1}\big(-s_I+t_I\big),&(2.21b)\cr}
$$
In terms of $s_I$, $t_I$, the scalar $\pi$--$b$ sector of $I_{[2]}$
takes the simple form
$$
\eqalignno{
I_{[2]}&={1\over 4\kappa^2}(\hbox{$1\over 18$}e^2)^{-{7\over 2}}
\int_{{\rm AdS}_4}d^4y (-\bar g_4)^{1\over 2}\sum_I z^{(7)}(k)&\cr
&\Big\{\hbox{$1\over 2$}\bar\nabla^\alpha \phi_{I\alpha\gamma}\bar\nabla^\beta
\phi_{I\beta}{}^\gamma\vphantom{\Big\{}
-\hbox{$1\over 2$}\bar\nabla^\alpha \phi_{I\alpha\beta}\bar\nabla^\beta 
\phi_I{}^\gamma{}_\gamma
+\hbox{$1\over 4$}\bar\nabla^\alpha \phi_I{}^\beta{}_\beta\bar\nabla_\alpha 
\phi_I{}^\gamma{}_\gamma
-\hbox{$1\over 4$}\bar\nabla^\alpha \phi_I{}^{\beta\gamma}\bar\nabla_\alpha 
\phi_{I\beta\gamma}\vphantom{\Big\{}&\cr
&+\hbox{$1\over 4$}m_{\phi I}'{}^2
\phi_I{}^\alpha{}_\alpha \phi_I{}^\beta{}_\beta
-\hbox{$1\over 4$}m_{\phi I}''{}^2\phi_I{}^{\alpha\beta}\phi_{I\alpha\beta}
\vphantom{\Big\{}&\cr
&+A_{sI}\Big[-\hbox{$1\over 2$}\bar\nabla^\alpha s_I\bar\nabla_\alpha s_I
-\hbox{$1\over 2$}m_{sI}{}^2s_Is_I\Big]\vphantom{\Big\{}
+A_{tI}\Big[-\hbox{$1\over 2$}\bar\nabla^\alpha t_I\bar\nabla_\alpha t_I
-\hbox{$1\over 2$}m_{tI}{}^2t_It_I\Big]
\vphantom{\Big\{}&\cr
&+u_Ip_I\Big[\phi_I{}^\alpha{}_\alpha
+v_{sI}\Big(\bar\nabla^\alpha\bar\nabla_\alpha s_I-m_{sI}{}^2s_I\Big)
+v_{tI}\Big(\bar\nabla^\alpha\bar\nabla_\alpha t_I-m_{tI}{}^2t_I\Big)\Big]
\vphantom{\Big\{}&\cr
&-w_Il_I{}^\alpha\Big[\bar\nabla^\beta\phi_{I\beta\alpha}
-\bar\nabla_\alpha\phi_I{}^\beta{}_\beta\Big]+\ldots\Big\},
&(2.22)\cr}
$$
where
$$\eqalignno
{A_{sI}&={2(k+3)k(k-1)\over (k+2)},&(2.23a)\cr
A_{tI}&={2(k+3)(k+6)(k+7)\over (k+4)},&(2.23b)\cr
m_{\phi I}'{}^2&=\hbox{$1\over 18$}e^2(k^2+6k+4),\vphantom{9\over 8}&(2.24a)\cr
m_{\phi I}''{}^2&=\hbox{$1\over 18$}e^2(k^2+6k-8),\vphantom{9\over 8}&(2.24b)
\cr
m_{sI}{}^2&=\hbox{$1\over 18$}e^2k(k-6),\vphantom{9\over 8}&(2.25a)\cr
m_{tI}{}^2&=\hbox{$1\over 18$}e^2(k+6)(k+12),\vphantom{9\over 8}&(2.25b)\cr
u_I&=(\hbox{$1\over 18$}e^2)^2\hbox{$3\over 7$}(k-1)k(k+6)(k+7),
\vphantom{9\over 8}&(2.26)\cr
v_{sI}&=(\hbox{$1\over 18$}e^2)^{-1}{1\over (k+2)},&(2.27a)\cr
v_{tI}&=(\hbox{$1\over 18$}e^2)^{-1}{1\over (k+4)},&(2.27b)\cr
w_I&=\hbox{$1\over 18$}e^2k(k+6).&(2.28)\cr}
$$
Note that, also in this case, the modes $s_I$ with $k=0,1$ do not appear 
so that these are gauge degrees of freedom.  
Fixing the gauge involves setting $p_I=0$, $l_{I\alpha}=0$. Thus, after gauge 
fixing, $s_I$, $t_I$ are free fields. On the $s_I$, $t_I$ mass shell
the constraints are again of the form (1.24a)--(1.24b).
The mass spectrum coincides with that found in [17], as expected.


\par\vskip.6cm
\item{\bf 3.} {\bf Conclusions and outlook}
\vskip.4cm
\par

We have determined from the 11d supergravity action compactified on
${\rm AdS}_7\times {\rm S}_4$  and 
${\rm AdS}_4\times {\rm S}_7$ the quadratic AdS 
bulk action for two Kaluza--Klein towers of scalar excitations 
described by the fields $s_I$, $t_I$.
In the AdS/CFT correspondence these fields are sources for the
boundary SCFT operators whose dimensions have been 
worked out in [9--12] from the knowledge of their
masses reported in [15--17].
Apart from reobtaining those masses, we have computed
the normalization factors $A_{sI}$, $A_{tI}$
which fix the normalization of the 2 point functions.
We have obtained this result
by isolating the leading quadratic term of the
relevant actions through field redefinitions.
This has allowed us to remain off--shell for 
the $s$, $t$ scalar fluctuations, thus avoiding the delicate 
task of keeping track of all possible boundary terms. On the other hand,
we have found that the on--shell method proposed in [4]
is not complete and does not yield the correct result,
as noticed also in [13].
Presumably certain boundary terms have been missed in [4].

In the ${\rm AdS}_7\times {\rm S}_4$ case, the coefficient 
$A_{sI}$ was also recently obtained in [13] using a different method
involving the calculation of certain three point couplings.
We agree with their result.

In the ${\rm AdS}_4\times {\rm S}_7$ compactification, the identification
of the normalizations is more laborious.  In this case the 
standard 3--form formulation of 11d supergravity gives
the off-shell description of a linear combination of 
the  scalars $s$, $t$ in terms of a vector fields $b_\alpha$. 
Therefore, we have used a dual description which produces directly both 
scalars and computed the relevant normalizations.

With these normalizations at hand, one can proceed to compute
the cubic $s$--$t$ self-coupling to obtain
the large $N$ limit of the three point correlation functions of 
the corresponding operators belonging to chiral representations of 
the maximally supersymmetric
$SCFT_3$ and $SCFT_6$ theories, thus adding extra pieces of information 
on these somewhat mysterious theories.

\vfill\eject

\vskip.6cm
\par\noindent
{\bf Acknowledgements.}  We are indebt to M. Tonin for 
providing illuminating comments and relevant literature.
FB would like to thank D. Z. Freedman and A. Hanany 
for helpful discussions and the INFN-MIT ``Bruno Rossi'' exchange program
for supporting his visit at MIT.

\vskip.6cm
\centerline{\bf REFERENCES}

\item{[1]} J. Maldacena, 
``The Large $N$ limit of Superconformal Field Theories and Supergravity'', 
Adv. Theor. Math. Phys. {\bf 2} (1998), 231, hep-th/9711200.

\item{[2]} S. S. Gubser, I. R. Klebanov and A. M. Polyakov, ``Gauge Theory
Correlators from Non-critical String Theory,'' 
Phys. Lett. {\bf B428} (1998), 105, hep-th/9802109.

\item{[3]} E. Witten, ``Anti-de Sitter Space and Holography'',
Adv. Theor. Math. Phys. {\bf 2} (1998), 253, hep-th/9802150.

\item{[4]}  S. Lee, S. Minwalla, M. Rangamani and N. Seiberg,
``Three Point Functions of Chiral Primary Operators in $D=4$, $N=4$ SYM 
at Large $N$'', Adv. Theor. Math. Phys. {\bf 2} (1998), 697, 
hep-th/9806074.

\item{[5]} D. Z. Freedman, S. D. Mathur, A. Matusis and L. Rastelli, 
as discussed in Freedman's conference lecture at ``Strings 98'', available 
at http://www.itp.ucsb.edu/string98/.

\item{[6]} H. Liu and A. A. Tseytlin, ``On Four Point Functions in the 
CFT/AdS Correspondence'',
Phys. Rev. {\bf D59} (1999) 086002, hep-th/9807097.

\item{[7]} D. Z. Freedman, S. D. Mathur, A. Matusis and L. Rastelli, 
``Comment on 4-Point Functions in the CFT/AdS Correspondence'',
hep-th/9808006.

\item{[8]}  E. D'Hoker and D.Z. Freedman
``General scalar exchange in $AdS_{d+1}$'',
hep-th/9811257.

\item{[9]} O. Aharony, Y. Oz and Z. Yin, 
``M Theory on ${\rm AdS}_p\times {\rm S}_{11-p}$ and Superconformal Field
Theories'', 
Phys. Lett. {\bf B430} (1998), 87, hep-th/9803051.

\item{[10]} S. Minwalla, ``Particles on {$AdS_{4/7}$} and Primary Operators 
on {$M_{2/5}$}-Brane World Volumes'', 
J. High Energy Phys. {\bf 9810} (1998) 002, hep-th/9803053.

\item{[11]} R.G. Leigh and M. Rozali, 
``The Large  $N$  Limit of the $(2,0)$ Superconformal Field Theory'', 
Phys. Lett. {\bf B431} (1998) 311, hep-th/9803068.

\item{[12]} E. Halyo, 
``Supergravity on {$AdS_{4/7}\times S^{7/4}$} and  M  Branes'', 
J. High Energy Phys. {\bf 04} (1998) 011, hep-th/9803077.

\item{[13]} R. Corrado, B. Florea and R. McNees, 
``Correlation Functions of Operators and Wilson Surfaces
in the $d=6$, $(0,2)$ Theory in the Large $N$ Limit'',
hep-th/9902153.

 \item{[14]} G. Arutyunov and S. Frolov, ``Quadratic Action for type IIB
Supergravity on ${\rm AdS}_5\times {\rm S}^5$'', hep-th/9811106.

\item{[15]} P. van Nieuwenhuizen,
``The Complete Mass Spectrum of $d=11$ Supergravity
Compactified on $S^4$ and a General Mass Formula 
for Arbitrary Cosets $M_4$'', Class. Quantum Grav. {\bf 2} (1985) 1.

\item{[16]}  B. Biran, A. Casher, F. Englert, M. Rooman and P. Spindel,
``The Fluctuating Seven Sphere in Eleven Dimensional Supergravity'',
Phys. Lett. {\bf 134B} (1984) 179.

\item{[17]} L. Castellani, R. D'Auria, P. Fr\'e, K. Pilch and 
P. van Nieuwenhuizen, 
``The bosonic Mass Formula for Freund--Rubin Solutions of
$d=11$ Supergravity on General Coset Manifolds'', 
Class. Quantum Grav. {\bf 1} (1984), 339.

\item{[18]} P. G. O. Freund and M. Rubin, 
``Dynamics of Dimensional Reduction'',
Phys. Lett. {\bf B97} (1980), 233.

\item{[19]} S. P. de Alwis,
``Coupling of Branes and Normalization of Effective
Actions in String / M Theory'',
Phys. Rev. {\bf D56} (1997), 7963, hep-th/9705139.

\item{[20]} I. Bandos, N. Berkovits and D. Sorokin,
``Duality Symmetric Eleven-Dimensional Supergravity and
its Coupling to M-Branes'',
Nucl. Phys. {\bf B522} (1998), 214, hep-th/9711055. 

\bye